\documentclass[twocolumn,floatfix,showpacs,aps,nofootinbib, showkeys]{revtex4}

\usepackage[dvips]{epsfig}
\usepackage[english]{babel}
\usepackage{bbm}
\usepackage{verbatim}
\usepackage{array}
\usepackage{amsmath}
\usepackage{multirow}
\usepackage{amsfonts}
\usepackage{amssymb}
\usepackage{hyperref}
\usepackage{leading}
\usepackage[dvipsnames]{xcolor}
\hypersetup{colorlinks=true,linkbordercolor=Blue,linkcolor=Blue, citecolor=Blue}
\usepackage{subeqnarray}
\usepackage{setspace}
\usepackage{indentfirst} 

\usepackage{gensymb}

\begin{document}

\title{Some Remarks on dual helicity flag-dipole spinors}

\author{R. J. Bueno Rogerio$^{1}$} \email{rodolforogerio@unifei.edu.br}
\author{C. H. Coronado Villalobos$^{2}$} \email{carlos.coronado@inpe.br}
\affiliation{$^{1}$Instituto de F\'isica e Qu\'imica, Universidade Federal de Itajub\'a - IFQ/UNIFEI, \\
Av. BPS 1303, CEP 37500-903, Itajub\'a - MG, Brasil.}
\affiliation{$^{2}$Instituto Nacional de Pesquisas Espaciais (INPE),\\
12227-010, S\~ao Jos\'e dos Campos, SP, Brazil.}


\begin{abstract}\noindent\rule{14.15cm}{0.5pt}\\
{\textbf{Abstract.}}
In this report we advance into a mapping procedure transmuting a single helicity spinor to a dual helicity spinor. Such a mathematical mechanism reveal us a class of spinor which fits into fourth class within Lounesto classification \cite{lounestolivro}. The focus of the present manuscript is to expose the algebraic construction of flag-dipole spinors and briefly explore its underlying physical and mathematical contents.
\\
\noindent\rule{14.15cm}{0.5pt}
\end{abstract}


\maketitle

\section{Introduction}\label{intro}
Spinors may be defined in several different fashions. In the context of Clifford algebra, the spinors are defined to be
elements of a left minimal ideal, whereas in the context of group theory we say that the spinors are carriers of the fundamental representation of the group \cite{beyondlounesto}. Physically, a spinor represents the quantum wave function of a spin-$1/2$ particle \cite{roldaorevisiting}, and as such the physical observables are obtained from real quadratic functional of these fields \cite{crawford1}. According to \cite{carmeli}, spinors are used extensively in physics; it is widely, accepted that they are more fundamental than tensors (when the spacetime itself is represented by a manifold endowed with a Riemannian - or Lorenzian - metric structure) and the easy way to see this fact is the results obtained in general relativity theory by using spinors, results that could not have been obtained by using tensor methods only \cite{introducingspinors}.

The Dirac equation is one of the most impressive successes in all of physics conceived from the purely theoretical reason to be a covariant first-order derivative field equation, it turned out to account for spin and matter/antimatter duality \cite{fabbrigeneral}. It is known that the full Lorentz group is composed by rotation generators ($\vec{J}$), boost generators ($\vec{K}$) and discrete symmetries: parity ($P$) and time reversion ($T$) \cite{weinberg1}, we also have the charge conjugation symmetry ($C$), nevertheless, it does not stand for a Lorentz transformation. 

Dirac spinors, which are responsible to describe the electron and its corresponding antiparticle the positron, are built within the full Lorentz group, keeping, usually, as the transformation law between the spinorial components, left-hand and right-hand components, the parity symmetry. In other words, this symmetry is exactly the link element between both parts of the representation space, i.e., it connects the subspaces $(j,0)$ and $(0,j)$ \cite{diracpauli}. 

In reference \cite{speranca}, the author derive in a new fashion the parity operator ($P$), in the spin $1/2$ representation, thus, it reads
\begin{equation}
P = m^{-1}\gamma_{\mu}p^{\mu},
\end{equation} 
which allows us to evoke the Dirac equation from the space-time symmetries alone. For Dirac spinors parity is used, and as an inexorable and direct consequence the Dirac dynamics is reached. This relation is literal: in acting on spinors, the parity operator is the Dirac operator and vice-versa. As a consequence, being the different sectors of the representation space related by means of another procedure (where parity plays no role), the Dirac dynamics is no longer expected \cite{elkograviton}.
Nevertheless, there are cases where the representation spaces are not connected by discrete symmetries, like in the Elko spinors case, proposed in reference \cite{jcap}. In this case the transformation law between the spinorial still related by the Wigner Time-Reversal operator, $\Theta$, such observation was first made by Ramond in \cite{ramond}, in such a way, that the above mentioned spinors do not belong to the Full Lorentz group any more, belonging, then, to the proper orthochronous Lorentz subgroup, denoted by $\mathcal{L}^{\uparrow}_{+}$. 

According to the Lounesto classification there are six disjoint classes of spinors \cite{lounestolivro}. The first three classes stand for regular spinors fields for spin $1/2$ fermions (covering three types of Dirac spinors) and the remaining three classes stand for singular spinors, encompassing type-4 (flag-dipole), Majorana (flag-pole) and Weyl spinors fields.

Up to the present moment, all the above mentioned spinor fields (and also the associated quantum field operators) are well established, except, then, for the type-4 spinor \cite{chengflagdipole}. Type-4 spinors composes a set of very rare spinors in the literature. Their first appearance, in a very specific scenario, can be seen in \cite{esk} and posteriorly \cite{roldaomeert}, these kind of singular spinors work appropriately as a source of matter in ESK gravity, being the first explicit construction in Physics of flag-dipole spinors \cite{roldaonewspinor}. The aforementioned spinor fields were shown to be solution of the Dirac field equation in a torsional framework in \cite{fabbrifr, vignolofr}.
Although there is no quantum field operator constructed based on type-4 spinors, it is expected that it does not respect the full Lorentz symmetries. This spinors are potential candidates to describe dark matter, dark energy and to construct mass-dimension-one fermions, as remarked in \cite{cavalcanti4}. To the best of our knowledge, there is compelling evidence in astrophysics and cosmology that most of the mass of the Universe is composed of a new form of non baryonic dark matter, as it can be seen in \cite{Pereira2019, pereira2018evolution, pereira2017lambda}, there is a lack of evidence of the existence of new physics at LHC (Large Hadron Collider) and other particle physics experiments. On the theory side, many specific models with new particles and interactions beyond the standard model have been proposed to account for dark matter \cite{mirco}. 

The paper is organized as follows: Sect.\ref{conceitos} we briefly expose some basic conceptions about spinors transformation rules. Sect.\ref{mapeamento}, we provide the main details of the programme of mapping single helicity spinors into dual helicity spinors, highlighting the main output that we reach. Then, in Sect.\ref{dinamicacpt}, we weave some comments about type-4 spinors behaviour under Dirac and also parity operator, we show that the introduced spinors are not ruled by the Dirac equation but rather the Klein-Gordon equation. Such results provides a hint towards mass dimensionality of the spinors at hand. Based on the Wigner classification, we also focus in establish some details about $C$, $P$ and $T$ discrete symmetries and also the observance of $(CPT)^{2}$. Finally, in Sect.\ref{remarks} we conclude and present some outlooks. 

\section{Foreword: Main concepts and setting up the notation}\label{conceitos}
To obtain an explicit form of a given $\psi(\boldsymbol{p})$ spinor we call for the rest spinors, $\psi(\boldsymbol{0})$. For an arbitrary momentum, we have the following condition
\begin{equation}\label{1}
\psi(\boldsymbol{p}) = e^{i\kappa.\varphi}\psi(\boldsymbol{0}),
\end{equation}
where the $\psi(\boldsymbol{0})$ is a direct sum of the $(1/2, 0)$ and $(0, 1/2)$ Weyl spinors spinor, defined as 
\begin{equation}
\psi(\boldsymbol{0}) = \left(\begin{array}{c}
\phi_R(\boldsymbol{0}) \\ 
\phi_L(\boldsymbol{0})
\end{array} \right).
\end{equation}
Thus, in spherically coordinate system the right-hand and left-hand components, in the rest-frame referential are defined as 
\begin{equation}
\phi_R^{+}(\boldsymbol{0}) = \phi_L^{+}(\boldsymbol{0}) = \sqrt{m}\left(\begin{array}{c}
\cos(\theta/2)e^{-i\phi/2} \\ 
\sin(\theta/2)e^{i\phi/2}
\end{array}\right), 
\end{equation} 
and 
\begin{equation}
\phi_R^{-}(\boldsymbol{0}) = \phi_L^{-}(\boldsymbol{0}) = \sqrt{m}\left(\begin{array}{c}
-\sin(\theta/2)e^{-i\phi/2} \\ 
\cos(\theta/2)e^{i\phi/2}
\end{array}\right).
\end{equation} 
the reader is cautioned to do not naively neglect the relation $\phi_R^{\pm}(\boldsymbol{0}) = -\phi_L^{\pm}(\boldsymbol{0})$ when dealing antiparticles, as highlighted in \cite{nogo, nondirac}. An inspection of the helicity operator furnish
\begin{equation}
\vec{\sigma}\cdot\hat{p}\; \phi^{\pm}(\boldsymbol{0}) = \pm \phi^{\pm}(\boldsymbol{0}).
\end{equation}
Thus, the boost operator is defined as follows
\begin{eqnarray}\label{boostoperator}
e^{i\kappa.\varphi} = \sqrt{\frac{E+m}{2m}}\left(\begin{array}{cc}
\mathbbm{1}+\frac{\vec{\sigma}.\hat{p}}{E+m} & 0 \\ 
0 & \mathbbm{1}-\frac{\vec{\sigma}.\hat{p}}{E+m}
\end{array} \right),
\end{eqnarray}
such programme allow one to write a boosted spinor in terms of a rest frame spinor, as we started our explanation in \eqref{1}.

\section{Mapping the spinors: Introducing type-4 spinors}\label{mapeamento}
This section is reserved for a temptation on a mapping procedure between single helity spinors into dual helicity spinors. The path to be followed here, is based on defining a mapping matrix ($M$) responsible to transmute a single helicity spinor into a dual helicity one. To start this task, we employ the following relations 
\begin{eqnarray}
M_{1}\psi_{\{+,+\}}= \Lambda_{\{+,-\}}, \label{lamb1}
\\
M_{2}\psi_{\{+,+\}}= \Lambda_{\{-,+\}}, \label{lamb2}
\\
M_{3}\psi_{\{-,-\}}= \Lambda_{\{+,-\}}, \label{lamb3}
\\
M_{4}\psi_{\{-,-\}}= \Lambda_{\{-,+\}}, \label{lamb4}
\end{eqnarray}
where the \emph{single helicity} spinors $\psi_{\{+,+\}}$ and $\psi_{\{-,-\}}$ are displayed in \cite{diracpauli,nondirac}, and the $M$ matrix commutes with the boost operators given in \eqref{boostoperator}.
Please, note that the relations \eqref{lamb1} and \eqref{lamb3} provide the same spinor $\Lambda$ (unless a global minus sign) and the same can be inferred to \eqref{lamb2} and \eqref{lamb4}. The mapping task is accomplished imposing the following relations, first imposing to a component which carry positive eigenvalue transmute to a component which carry negative eigenvalue, given task is accomplished by defining the $\mathcal{O}$ operator
\begin{eqnarray}
\vec{\sigma}\cdot\hat{p}\;\mathcal{O}_{(+-)}\phi^{+} = -\phi^{-},
\end{eqnarray}
finally, we impose the opposite
\begin{eqnarray}
\vec{\sigma}\cdot\hat{p}\;\mathcal{O}_{(-+)}\phi^{-} = +\phi^{+}.
\end{eqnarray}
Where the $\mathcal{O}$ operator stand for a $2\times 2$ matrix which changes the component's helicity. Note that the lower index ($+,-$) and ($-,+$) indicates that the operator $\mathcal{O}$ takes a component which carry positive eigenvalue, to one which carry negative one and \emph{vice-versa}. Such operator composes a more general operator $M$, which acts over the spinors, changing its component helicity. 
The $M$ matrix can be displayed as follows
\begin{equation}
M^{L} = \left(\begin{array}{cc}
\mathbbm{1} & 0 \\ 
0 & \mathcal{O}
\end{array} \right),
\end{equation}
which keeps the right-hand helicity unchanged and acts only on the right-hand component, and we also may define
\begin{equation}
M^{R} = \left(\begin{array}{cc}
\mathcal{O} & 0 \\ 
0 & \mathbbm{1}
\end{array} \right),
\end{equation}
where it acts only on the right-hand component, where $\mathbbm{1}$ stand for $2\times 2$ identity matrix.
After a direct calculation,  given matrix is explicitly given by
\begin{equation}
M^{L}_{(+-)} = \left(\begin{array}{cccc}
1 & 0 & 0 & 0 \\ 
0 & 1 & 0 & 0 \\ 
0 & 0 & 0 & -e^{-i\phi} \\ 
0 & 0 & e^{i\phi} & 0
\end{array}\right),
\end{equation}
we also may construct
\begin{equation}
 M^{R}_{(+-)} = \left(\begin{array}{cccc}
0 & -e^{-i\phi} & 0 & 0 \\ 
e^{i\phi} & 0 & 0 & 0 \\ 
0 & 0 & 1 & 0 \\ 
0 & 0 & 0 & 1
\end{array}\right),
\end{equation}
another possibility is given by
\begin{equation}
M^{L}_{(-+)} = \left(\begin{array}{cccc}
1 & 0 & 0 & 0 \\ 
0 & 1 & 0 & 0 \\ 
0 & 0 & 0 & e^{-i\phi} \\ 
0 & 0 & -e^{i\phi} & 0
\end{array}\right), 
\end{equation}
and 
\begin{equation}
 M^{R}_{(-+)} = \left(\begin{array}{cccc}
0 & e^{-i\phi} & 0 & 0 \\ 
-e^{i\phi} & 0 & 0 & 0 \\ 
0 & 0 & 1 & 0 \\ 
0 & 0 & 0 & 1
\end{array}\right),
\end{equation}
where the upper index $R$ and $L$ stand for the component (right-hand or left-hand) where $M$ acts, all the above $M$ matrix hold inverse, ensuring the possibility of an invertible mapping procedure. With the spinor given in \cite{nondirac} at hand, we can explicit built the following set of dual helicity spinors in the rest frame referential
\begin{equation}
\Lambda_{\{+,-\}}(\boldsymbol{0})=\sqrt{m}\left(\begin{array}{c}
\alpha_{+}\cos(\theta/2)e^{-i\phi/2}\\
\alpha_{+}\sin(\theta/2)e^{i\phi/2}\\
-\beta_{+}\sin(\theta/2)e^{-i\phi/2}\\
\beta_{+}\cos(\theta/2)e^{i\phi/2}\\
\end{array}
\right),
\end{equation}
and
\begin{equation}
\Lambda_{\{-,+\}}(\boldsymbol{0})=\sqrt{m}\left(\begin{array}{c}
-\alpha_{-}\sin(\theta/2)e^{-i\phi/2}\\
\alpha_{-}\cos(\theta/2)e^{i\phi/2}\\
\beta_{-}\cos(\theta/2)e^{-i\phi/2}\\
\beta_{-}\sin(\theta/2)e^{i\phi/2}
\end{array}
\right),
\end{equation}
where the parameters $\alpha$ and $\beta$ are constant phases to be further determined. Note that the $(0, 1/2)$ and $(1/2,0)$ representation spaces are now connected by the Wigner Time-Reversal operator, $\Theta$, similarly as Elko do \cite{aaca}. Taking advantage of the last above feature and then taking a rest frame spinor to a momentum arbitrary referential, using the boost operator presented in \eqref{boostoperator}, it follows
\begin{equation}
\Lambda_{\{\pm,\mp\}}(\boldsymbol{p}) =  e^{i\kappa.\varphi}\Lambda_{\{\pm,\mp\}}(\boldsymbol{0}),
\end{equation}
then, such relation allow one to write\footnote{In order to make the notation clear, we have defined the Lorentz boost parameters as $\mathcal{B}_{\pm}\equiv\sqrt{\frac{E+m}{2m}}\big(1\pm \frac{p}{E+m}\big)$.} 
\begin{equation}\label{spinor1}
\Lambda_{\{+,-\}}(\boldsymbol{p})=\mathcal{B}_{+}\left(\begin{array}{c}
-\alpha_{+}\Theta\phi_L^{-*}(\boldsymbol{0}) \\ 
\beta_+\phi_L^{-}(\boldsymbol{0})
\end{array} \right),
\end{equation}
and
\begin{equation}\label{spinor2}
 \Lambda_{\{-,+\}}(\boldsymbol{p})=\mathcal{B}_{-}\left(\begin{array}{c}
\alpha_{-}\Theta\phi_L^{+*}(\boldsymbol{0}) \\ 
\beta_{-}\phi_L^{+}(\boldsymbol{0})
\end{array} \right),
\end{equation}
given spinors fits into fourth class within Lounesto classification \cite{lounestolivro} if one constrain the parameters $\vert\alpha\vert^{2}\neq\vert\beta\vert^{2}$. In such a way, we have explicitly defined a type-4 spinor field. 
A quick inspection on the flag-dipole spinors relations, presented in Ref \cite{cavalcanticlassification}, reveal us that is also possible to type-4 spinors carrying dual helicity feature, so, on this essay what we have is a matter standing for a particular case of the two presented in the above Reference.
A parenthetic remark, for the Elko spinors the modulus of the phases are equal, leading then, such spinors to fit fifth class\footnote{The crucial difference between fourth and fifth Lounesto classes, reside in the axial-vector, the bilinear $\boldsymbol{K}=K_{\mu}\gamma^{\mu}$, standing non-null (null) for the first case (second case).} within Lounesto classification \cite{whereareelko, hoffdirac}, thus, the only way to reconstruct Elko spinors from \eqref{spinor1} and \eqref{spinor2}, is accomplished by a convenient choice of phases. Remarkably enough, Elko spinors have the following constraint of phases $\alpha=\pm i$ and $\beta=1$, holding the relation $\vert\alpha\vert^{2}=\vert\beta\vert^{2}$, thus, if a same akin reasoning is taking into account, type-4 spinors obey the charge-conjugation relation $C\Lambda = \pm\Lambda$.   

\section{Some comments related with dynamic and discrete symmetries}\label{dinamicacpt}
Regarding to the current literature about spinors, it is stated, and first stressed by Lounesto, that type-4 spinors can not satisfy the Dirac equation \cite{lounestolivro}. Guided by the Lounesto statement, the first point that we should emphasize concerns to the connection between the representation spaces, for type-4 spinors we found that $(0, 1/2)$ and $(1/2, 0)$ representation spaces are connected by the Wigner time-reversal operator whereas for the Dirac case such connection occurs via parity relation \cite{ryder, diracpauli}. However, as expected for all physical fields, type-4 spinors obey the Klein-Gordon equation. As defined in \cite{speranca}, we invoke the parity operator, $P=m^{-1}\gamma_{\mu}p^{\mu}$, looking towards obtain some information about the related dynamic, thus, 
\begin{equation}\label{dirac}
\gamma_{\mu}p^{\mu}\Lambda(\boldsymbol{p}) \neq m\Lambda(\boldsymbol{p}),
\end{equation}
such unambiguous calculation, presented in relation \eqref{dirac}, hold for both spinors displayed in \eqref{spinor1} and \eqref{spinor2}, i.e., the momentum space Dirac operator does not annihilate the introduced spinors (or any dual helicity spinor), without any phase constraint between $\alpha_{\pm}$ and $\beta_{\pm}$, one can not stablish any relation between the flag-dipole spinors presented above. Thus, such a direct observance provide a concrete conclusion about the Lounesto's previous observations about type-4 spinors dynamic, as presented in his preprint \cite{lounestolivro}. The aforementioned remark is a hint towards the canonical mass dimension related to the type-4 spinors, which allow us to assert that it is not $3/2$, as conventionally was expected for fermions. Acting again with the Dirac operator $\gamma_{\mu}p^{\mu}$, in \eqref{dirac}, it yields
\begin{equation}
(\Box +m^2)\Lambda(\boldsymbol{p}) = 0,
\end{equation}
holding the Klein-Gordon equation. 
Regarding to the discrete symmetries we have the following results $C^{2}=\mathbbm{1}$, $P^{2}=\mathbbm{1}$ and $T^{2}=-\mathbbm{1}$, as expected for the fermions which belong to the Standard Model \cite{Wigner1}, while the unconventional output comes from $(CPT)^2=+\mathbbm{1}$ rather than $(CPT)^2=-\mathbbm{1}$, nevertheless, corresponding to the third Wigner class of spin one-half particles \cite{Wigner1}.  

It is worth to emphasize that up to now the introduced spinors are not eigenspinors of parity and neither charge conjugation operator. In this vein, we are not able to defined which spinor describe particle or antiparticle, due to the aforementioned reason, it precludes us to proceed with the computation of the spin sums and the quantum field operator. Guided by the dynamic that such spinors are ruled and also due to the dual helicity feature, in a speculative way, we may say that the associated quantum field operator may hold similarity with Elko spinor fields.

We restrict ourselves in the present essay addressing essentially the algebraic construction of the flag-dipole spinors, confining our attention to the rudimentary aspects of it, looking towards the accordance with what was stated in the  previous literature. Perhaps, this the first time that it explicitly emerges in the literature. Regarding to phenomenological implications, once such spinors do not hold charge conjugacy feature, we may expect that it does not hold suppressed interaction with electromagnetic field, as example, however, at this very moment it is too early to advance into such assumption. Details concerning other subjects, e.g., couplings and interactions, cosmological implications, definition of an appropriated quantum field operator and \emph{etc}, deserves our efforts to be soon developed. 
 
\section{Final Remarks}\label{remarks}
The direct observance of type-4 (flag-dipole) fermions in physics is extremely rare. Besides the previous examples, cited in the scope of this manuscript, that arise in the literature as solutions of the derived equations of motion of specific frameworks, here we approach them by a different path; they are solutions of a given mathematical mapping procedure among single helicity and dual helicity spinors. 
All the solutions here developed are extremely important from a theoretical point of view, due to the fact that they evince that given spinor fields are not necessarily determined by their dynamics, however, a detailed discussion on their  algebraic properties should be stressed.

In the present theoretical manuscript, we (unveiled) reached to an explicit set of dual helicity spinors, which representation spaces $(0,1/2)$ and $(1/2,0)$ are connected by the Wigner Time-Reversal operator rather than parity symmetry playing such role and it fits into the fourth class within Lounesto classification. Such result was possible by means a mathematical device which maps single helicity spinors into dual helicity spinors. To ensure the consistency of the results, we stated that under certain phases fixation, it becomes possible to recover the well-known Elko spinors, as a very particular case. 

We also showed that type-4 spinors do not have an specific dynamic, and, as presented, the Dirac operator does not annihilate them. Thus,  the only wave equation that type-4 spinors obey is the Klein-Gordon equation, as it is expected for all the physical fields. As a consequence, due to the last aforementioned features, it reveals us that the mass dimension of the field can not be $3/2$ as for the Dirac fermions, however, the correct path to ascertain the mass dimensionality makes necessary to built the quantum field operators, promoting the spinors to play the role of expansion coefficients of such physical amount. 

We elucidate that it remains as open windows for further investigations, we look for define an adjoint spinor which provide a scalar, real and non-vanishing norm under Lorentz transformations, such a procedure, then, will allow us to look for an appropriate quantum field operator, a field propagator and to determine a positive definite Hamiltonian in terms of the creator and annihilator operators, yielding the Fermi-Dirac statistic for fermionic fields. 

\section{Acknowledgements}
RJBR thanks CNPq Grant N$^{\circ}$. 155675/2018-4 for the financial support and CHCV thanks CNPq PCI Grant N$^{\circ}$. 300236/2019-0 for the financial support.

\bibliographystyle{unsrt}
\bibliography{refs}

\begin{thebibliography}{10}

\bibitem{lounestolivro}
P.~Lounesto.
\newblock {\em Clifford algebras and spinors}, volume 286.
\newblock Cambridge university press, 2001.

\bibitem{beyondlounesto}
C.~H. Coronado~Villalobos, R.~J. Bueno~Rogerio, A.~R. Aguirre, and D.~Beghetto.
\newblock On the generalized spinor field classification: Beyond the {L}ounesto
  classification.
\newblock {\em arXiv:1906.11622 [hep-th]}, 2019.

\bibitem{roldaorevisiting}
R.~da~Rocha and J.~Vaz~Jr.
\newblock {R}evisiting {C}lifford algebras and spinors {II}: {W}eyl spinors in
  {C}l$(3,0)$ and {C}l$(0, 3)$ and the {D}irac equation.
\newblock {\em arXiv preprint math-ph/0412075}, 2004.

\bibitem{crawford1}
J.~P. Crawford.
\newblock Bispinor geometry for even-dimensional space-time.
\newblock {\em Journal of mathematical physics}, 31(8):1991--1997, 1990.

\bibitem{carmeli}
M.~Carmeli and S.~Malin.
\newblock {\em Theory of spinors: {A}n introduction}.
\newblock World Scientific Publishing Company, 2000.

\bibitem{introducingspinors}
A.~M. Steane.
\newblock {A}n introduction to spinors.
\newblock {\em arXiv preprint arXiv:1312.3824}, 2013.

\bibitem{fabbrigeneral}
L.~Fabbri.
\newblock General dynamics of spinors.
\newblock {\em Advances in Applied Clifford Algebras}, 27(4):2901--2920, 2017.

\bibitem{weinberg1}
S.~Weinberg.
\newblock {\em The quantum theory of fields}, volume~1.
\newblock Cambridge university press, 1995.

\bibitem{diracpauli}
C.~H. Coronado~Villalobos and R.~J. Bueno~Rogerio.
\newblock The connection between {D}irac dynamic and parity symmetry.
\newblock {\em EPL (Europhysics Letters)}, 116(6):60007, 2016.

\bibitem{speranca}
L.~D. Sperança.
\newblock An identification of the {D}irac operator with the parity operator.
\newblock {\em International Journal of Modern Physics D}, 23(14):1444003,
  2014.

\bibitem{elkograviton}
R.~J. Bueno~Rogerio, R.~de~C. Lima, L.~Duarte, J.~M. Hoff~da Silva, M.~Dias,
  and C.~R. Senise~Jr.
\newblock Mass dimension one fermions and their gravitational interaction.
\newblock {\em arXiv preprint arXiv:1902.01379}, 2019.

\bibitem{jcap}
D.~V. Ahluwalia-Khalilova and D.~Grumiller.
\newblock Spin-half fermions with mass dimension one: theory, phenomenology,
  and dark matter.
\newblock {\em Journal of Cosmology and Astroparticle Physics}, 2005(07):012,
  2005.

\bibitem{ramond}
P.~Ramond.
\newblock Field theory: {A} modern primer.
\newblock {\em Front. Phys.}, 74:1--397, 1981.

\bibitem{chengflagdipole}
C.~Y. Lee.
\newblock {Mass dimension one fermions from flag dipole spinors}.
\newblock 2018.

\bibitem{esk}
R.~da~Rocha, Luca Fabbri, J.~M. Hoff~da Silva, R.~T. Cavalcanti, and J.~A.
  Silva-Neto.
\newblock Flag-dipole spinor fields in {ESK} gravities.
\newblock {\em Journal of Mathematical Physics}, 54(10):102505, 2013.

\bibitem{roldaomeert}
P.~Meert and R.~da~Rocha.
\newblock The emergence of flagpole and flag-dipole fermions in fluid/gravity
  correspondence.
\newblock {\em The European Physical Journal C}, 78(12):1012, 2018.

\bibitem{roldaonewspinor}
R.~da~Rocha.
\newblock New spinor fields classes and applications.
\newblock {\em Journal of Physics: Conference Series}, 804:012012, Jan 2017.

\bibitem{fabbrifr}
L.~Fabbri and S.~Vignolo.
\newblock Dirac fields in f(r)-gravity with torsion.
\newblock {\em Classical and Quantum Gravity}, 28(12):125002, 2011.

\bibitem{vignolofr}
S.~Vignolo, L.~Fabbri, and R.~Cianci.
\newblock Dirac spinors in {B}ianchi-{I} $f(r)$-cosmology with torsion.
\newblock {\em Journal of Mathematical Physics}, 52(11):112502, 2011.

\bibitem{cavalcanti4}
R.~T. Cavalcanti.
\newblock Looking for the classification of singular spinor fields dynamics and
  other mass dimension one fermions: Characterization of spinor fields.
\newblock {\em International Journal of Modern Physics D}, 23, 07 2014.

\bibitem{Pereira2019}
S.~H. Pereira, M.~E.~S. Alves, and T.~M. Guimar{\~a}es.
\newblock An unified cosmological evolution driven by a mass dimension one
  fermionic field.
\newblock {\em The European Physical Journal C}, 79(6):543, Jun 2019.

\bibitem{pereira2018evolution}
S.~H. Pereira, R.~F.~L. Holanda, and A.~P.~S. Souza.
\newblock Evolution of the universe driven by a mass-dimension-one fermion
  field.
\newblock {\em EPL (Europhysics Letters)}, 120(3):31001, 2018.

\bibitem{pereira2017lambda}
S.H. Pereira, A.~P.~S. Souza, J.~M. Hoff~da Silva, and J.~F. Jesus.
\newblock $\lambda$ (t) cosmology induced by a slowly varying elko field.
\newblock {\em Journal of Cosmology and Astroparticle Physics}, 2017(01):055,
  2017.

\bibitem{mirco}
M.~Cannoni.
\newblock Relativistic and nonrelativistic annihilation of dark matter: a
  sanity check using an effective field theory approach.
\newblock {\em The European Physical Journal C}, 76(3):137, Mar 2016.

\bibitem{nogo}
D.~V Ahluwalia.
\newblock Evading {W}einberg's no-go theorem to construct mass dimension one
  fermions: Constructing darkness.
\newblock {\em EPL (Europhysics Letters)}, 118(6):60001, 2017.

\bibitem{nondirac}
R.~J. Bueno~Rogerio and C.~H. Coronado~Villalobos.
\newblock Non-standard {D}irac adjoint spinor: The emergence of a new dual.
\newblock {\em EPL (Europhysics Letters)}, 121(2):21001, 2018.

\bibitem{aaca}
D.~V. Ahluwalia.
\newblock The theory of local mass dimension one fermions of spin one half.
\newblock {\em Advances in Applied Clifford Algebras}, 27(3):2247--2285, Sep
  2017.

\bibitem{cavalcanticlassification}
R.~T. Cavalcanti.
\newblock Classification of singular spinor fields and other mass dimension one
  fermions.
\newblock {\em International Journal of Modern Physics D}, 23(14):1444002,
  2014.

\bibitem{whereareelko}
R.~da~Rocha and W.~A. Rodrigues~Jr.
\newblock Where are {ELKO} spinor fields in {L}ounesto spinor field
  classification?
\newblock {\em Modern Physics Letters A}, 21(01):65--74, 2006.

\bibitem{hoffdirac}
J.~M. Hoff~da Silva and R.~Da~Rocha.
\newblock From {D}irac action to {ELKO} action.
\newblock {\em International Journal of Modern Physics A},
  24(16n17):3227--3242, 2009.

\bibitem{ryder}
L.~H. Ryder.
\newblock {\em Quantum Field Theory}.
\newblock Cambridge University Press, 2 edition, 1996.

\bibitem{Wigner1}
E.~Wigner.
\newblock On unitary representations of the inhomogeneous {L}orentz group.
\newblock {\em Annals of Mathematics}, 40(1):149--204, 1939.

\end{thebibliography}

\end{document}